\documentclass[fleqn,10pt]{wlscirep}
\title{Traction force microscopy with optimized regularization and automated Bayesian parameter selection for comparing cells}

\author[1]{Yunfei Huang}
\author[2]{Christoph Schell}
\author[4,5,6]{Tobias B. Huber}
\author[1]{Ahmet Nihat Şimşek}
\author[3]{Nils Hersch}
\author[3]{Rudolf Merkel}
\author[1]{Gerhard Gompper}
\author[1,*]{Benedikt Sabass}
\affil[1]{Theoretical Soft Matter and Biophysics, Institute of Complex Systems-2 and Institute for Advanced Simulation, Forschungszentrum Juelich, D-52425, Juelich, Germany}
\affil[2]{Institut für Klinische Pathologie, Universitätsklinikum Freiburg, D-79002, Freiburg, Germany}
\affil[3]{Biomechanics, Institute of Complex Systems-7, Forschungszentrum Juelich, D-52425, Juelich, Germany}
\affil[4]{Department of Medicine IV, Faculty of Medicine, Medical Center – University of Freiburg, Germany}
\affil[5]{BIOSS Center for Biological Signalling Studies, Albert-Ludwigs-University Freiburg, Germany}
\affil[6]{III. Department of Medicine, University Medical Center Hamburg-Eppendorf, Hamburg, Germany}
\affil[*]{b.sabass@fz-juelich.de}

\keywords{Traction force microscopy, Inverse problems, Elastic net regularization, Bayesian regularization}

\begin{abstract}
Adherent cells exert traction forces on to their environment, which allows them to migrate, to maintain tissue integrity, and to form complex multicellular structures during developmental morphogenesis. Traction force microscopy (TFM) enables the measurement of traction forces on an elastic substrate and thereby provides quantitative information on cellular mechanics in a perturbation-free fashion. In TFM, traction is usually calculated via the solution of a linear system, which is complicated by undersampled input data, acquisition noise, and large condition numbers for some methods. Therefore, standard TFM algorithms either employ data filtering or regularization. However, these approaches require a manual selection of filter- or regularization parameters and consequently exhibit a substantial degree of subjectiveness. This shortcoming is particularly serious when cells in different conditions are to be compared because optimal noise suppression needs to be adapted for every situation, which invariably results in systematic errors.\\
Here, we systematically test the performance of new methods from computer vision and Bayesian inference for solving the inverse problem in TFM. We compare two classical schemes, L1- and L2-regularization, with three previously untested schemes, namely Elastic Net regularization, Proximal Gradient Lasso, and Proximal Gradient Elastic Net. Overall, we find that Elastic Net regularization, which combines L1 and L2 regularization, outperforms all other methods with regard to accuracy of traction reconstruction. Next, we develop two methods, Bayesian L2 regularization and Advanced Bayesian L2 regularization, for automatic, optimal L2 regularization. Using artificial data and experimental data, we show that these methods enable robust reconstruction of traction without requiring a difficult selection of regularization parameters specifically for each data set. Thus, Bayesian methods can mitigate the considerable uncertainty inherent in comparing cellular tractions in different conditions.
\end{abstract}

\begin{document}
\flushbottom
\maketitle

\thispagestyle{empty}

\section*{Introduction}
Mechanical forces between cells and their embedding matrix are essential for a variety of biological processes, ranging from migration of cells -- including immune cells and cancer cells -- to tissue maintenance and organ development, see\cite{Geiger2001,Sheetz2006,Hinz2007,Weaver2009,Reinhart-King2010,ringer2017multiplexing,lecuit2011force} for only a few of the many review articles on this topic. Many of the relevant processes occur on a micrometer, or sub-micrometer lengthscale, for instance in nascent cell adhesion sites, filopodia, and bacterial adhesion. To understand these processes' mechanics and its biological control, reliable and accurate methods for measurement of cellular forces are required.

\begin{figure}[ht]
	\centering
	\includegraphics[width=0.8\linewidth]{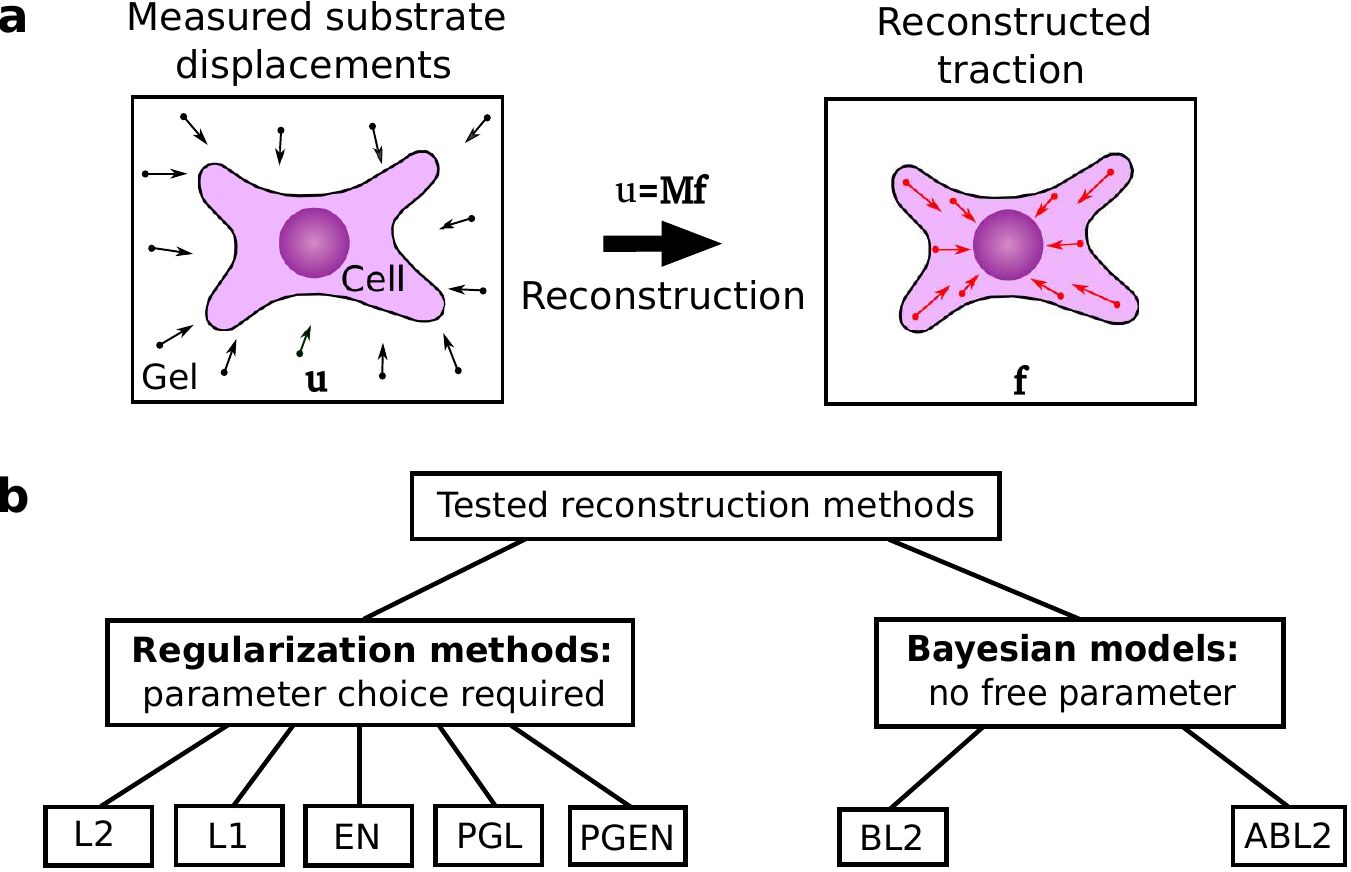}
	\caption{Schematic representation of a typical traction force microscopy (TFM) setup and different reconstruction methods for TFM. (a)~Cells are plated on a planar gel substrate containing fiducial markers. Tracking the markers allows to infer the deformations $\mathbf{u}$ in the surrounding of the cell. These deformations are linearly related to the cellular traction forces $\mathbf{f}$. The problem of calculating traction $\mathbf{f}$ from displacement $\mathbf{u}$ is associated with inverting an ill-conditioned matrix $\mathbf{M}$. This problem can be solved with different reconstruction methods.
	(b)~In this work, we test five regularization methods for traction reconstruction: L2 regularization (L2), L1 regularization (L1), EN regularization (EN),  Proximal Gradient Lasso (PGL) and Proximal Gradient Elastic Net (PGEN). Furthermore, we develop two Bayesian approaches that do not have any free parameters, namely Bayesian L2 regularization (BL2) and Advanced Bayesian L2 regularization (ABL2).}
	\label{fig:1}
\end{figure}

Traction force microscopy (TFM) is a versatile and perturbation-free method yielding a spatial image of substrate stress exerted by cells on relatively soft elastic gel substrates. This method has its origins in pioneering work by Harris et al.~\cite{harris1980silicone}, who employed flexible silicone substrates to investigate the mechanical forces that cells generate. Today, traction force microscopy has become a method that is routinely used in laboratories studying cell biology and soft matter physics around the world.\cite{roy_rajfur_pomorski_jacobson_2002,LangeFabry,style2014traction,schwarz2015traction,polacheck2016measuring,roca2017quantifying} Traction force microscopy includes three distinct procedures, illustrated in Fig.~\ref{fig:1}(a): (1)~Cells are plated on an elastic substrate containing fiducial markers allowing to quantify gel deformation visually, for instance fluorescent beads or quantum dots.\cite{dembo1999stresses,bergert2016confocal} The deformations caused by adherent cells are recorded by taking images of the gel before and after removing the cell. (2)~A discrete gel displacement field $\mathbf{u}$ is calculated by tracking the markers. The most common techniques for tracking are particle tracking velocimetry (PTV) and particle image velocimetry (PIV). (3)~Finally, the traction force field $\mathbf{f}$ is calculated from the displacement field $\mathbf{u}$ by making use of a mechanical model of the elastic substrate. A variety of methods exist for this purpose, including finite element methods,\cite{yang2006determining,hur2009live,tang2014novel,soine2016measuring,kulkarni2018traction} boundary element methods\cite{dembo1996imaging,dembo1999stresses,sabass2008high} and methods operating in Fourier space\cite{butler2002traction,sabass2008high,franck2011three,style2014traction,kulkarni2018traction}. Usually, calculation of traction from displacement requires either filtering or regularization approaches to reduce the effect of noise. The TFM methodology is limited by two common serious issues that introduce systematic errors. First, the resolution of the measured traction is usually not high enough to resolve processes at micrometer-sized cellular structures. Secondly, the most commonly used TFM algorithms require the user to choose a filter or a regularization parameter, which introduces a considerable degree of subjectivity regarding smoothness and magnitude of the resulting traction. In this article, we suggest methods for improving the state-of-the art with respect to these issues. 

In the standard TFM approach it is assumed that the substrate is a homogeneous, isotropic, and linear elastic half-space. The mechanical model relating a continuous displacement field $U_i(\mathbf{x})$ to the traction force field $F_j(\mathbf{x'})$ on a two-dimensional ($\mathbf{x}=(x_1,x_2)$) surface of the gel is expressed as the integral equation\cite{landau1986theory}
 \begin{equation}
 U_i(\mathbf{x})=\int_{\Omega} \sum_{j=1}^2 G_{ij}(\mathbf{x}-\mathbf{x'})F_j(\mathbf{x'})\,\mathrm{d}^2\mathbf{x'},
 \label{eq:1}
 \end{equation}
where $\Omega$ denotes the whole surface of the substrate. The integrand contains a Green's function $G_{ij}(\mathbf{x})=(1+\nu)/(\pi E)[(1-\nu)\delta_{ij}/r+\nu x_ix_j/r^3]$, and $E$ and $\nu$ represent the Young modulus and Poisson ratio, respectively. We also write $r=|\mathbf{x}|$ and $\delta_{ij}$ is the Kronecker delta function. Calculation of the traction $F_j$ requires inversion of Eq.~(\ref{eq:1}). A very popular and practical approach is to solve Eq.~\eqref{eq:1} in Fourier space.\cite{butler2002traction,tolic2002spatial,sabass2008high} With this approach, the inversion is often directly feasible if noise in the displacement data has been filtered prior to calculation of the traction. Optimal filtering, however, requires input of a prior-defined filter function that imposes a smoothness constraint on the calculated traction. Moreover, spatial clustering of traction into sparse regions is not conserved when switching from real space to Fourier space. To take advantage of the sparsity of traction patterns for better reconstruction, one can solve Eq.~\eqref{eq:1} in real space. Here, the integral in Eq.~\eqref{eq:1} can be converted into a matrix product by discretizing the traction field $F_j$ and interpolating it as a piecewise linear, continuous function using  pyramidal shape functions ${h(\mathbf{x})}$.\cite{sabass2008high, han2015traction} We write the two-dimensional, discrete displacement field as a $2m\times1$ vector $\mathbf{u}$, where $m$ is the number of discretization nodes. The discrete traction field $\mathbf{f}$ is a $2n\times1$ vector, where $n$ is the number of nodes at which traction is prescribed. Then, Eq.~\eqref{eq:1} becomes  
\begin{equation}
\mathbf{u}=\mathbf{M}\mathbf{f} + \mathbf{s},
\label{eq:2}
\end{equation}
where we also explicitly included the linear acquisition noise $\mathbf{s}$ that is present the experimental data. The matrix $\mathbf{M}$ represents the coefficients of a discretized integration. It can be calculated with different techniques. The required convolution of the Green's function with the shape functions ${h(\mathbf{x})}$ can be done by numerical integration.\cite{sabass2008high} However, the computational effort can be significantly reduced by using the convolution theorem in Fourier space on regular grids.\cite{han2015traction} In the supplementary S1, we show how this method for fast calculation of $\mathbf{M}$ can be extended to irregularly spaced measurements with the help of the shift theorem. Regardless of the way how $\mathbf{M}$ is calculated, the condition number of $\mathbf{M}$ is typically very large. This means that even the smallest noise $\mathbf{s}$ leads to very large errors if a direct inversion of Eq.~\eqref{eq:2} is attempted. A further characteristic of TFM is that the experimental data is always undersampled, which introduces a large degree of uncertainty. The challenge in traction force microscopy is to turn Eq.~(\ref{eq:2}) into a well-posed problem, while providing a traction field $\mathbf{f}$ with correct magnitude at high spatial resolution. The standard approaches employed in TFM are L1 or L2 regularization that penalize norms of the solution to render traction calculation a well-defined numerical problem.

However, linear problems involving large, ill-conditioned matrices occur commonly throughout engineering and physics. Consequently, a variety of solution methods exist that not yet been employed in the context of TFM. For example, the elastic net regularization. Literally, the elastic net regularization behaves like a stretchable fishing net that retains "all the big fish" while removing the small background signal.\cite{zou2005regularization} An alternative are proximal gradient methods, which usually operate in wavelet space and employ adaptive or non-adaptive thresholding of high spatial frequencies.\cite{combettes2011proximal,schmidt2011convergence} Proximal gradient methods are for instance used for reconstructing lost parts of an image,\cite{mosci2010solving,peyre2008non,fadili2011total} for analysis of MRI data,\cite{michel2011total} and for analysis of genomic data.\cite{hannum2013genome,sokolov2016pathway} All regularization methods have the weakness that they require selection of one or more constant regularization parameters to allow discrimination of noise and signal.

Bayesian statistics provides one solution to this problem. From the Bayesian point of view, regularization parameters can be seen of as random variables that are picked for every experimental sample from a prior distribution. Thus, one infers the most probable value of the regularization parameters in a conceptually similar way as one infers the solution.\cite{mackay1992bayesian} Such methods do not require the choice of a regularization parameter and are therefore potentially less prone to subjectiveness than the classical regularization. 
The conceptual framework of employing hierarchical priors for data regularization has been exploited for a large number of different applications, for instance in astrophysics,\cite{suyu2006bayesian,vegetti2009bayesian,ghosh2015bayesian}, machine learning\cite{tipping2001sparse}, mechanical structure monitoring\cite{huang2014robust}, face recognition\cite{qiao2010sparsity}, and radar imagery\cite{zhao2014autofocus}. Therefore, it is to be expected that Bayesian analysis can be of great use for traction force reconstruction.

In this work, we systematically compare a range of different approaches for solving Eq.~\eqref{eq:2}. Altogether, we study the performance of seven methods, illustrated as a schematic diagram in Fig.~\ref{fig:1}(b). First, we test various regularization methods. Among the regularization methods, we complement the classical TFM approaches, L1- and L2 regularization, with previously untested methods from computer vision, namely Elastic Net (EN) regularization, Proximal Gradient Lasso (PGL), and Proximal Gradient Elastic Net (PGEN). We find that the new EN regularization scheme has a substantially improved accuracy as compared to previous approaches but requires considerable extra computational cost. 
Secondly, we seek to establish Bayesian models that can automatically perform an optimal regularization of the data. Initial tests indicate that different freely available Bayesian hierarchical models are of little use for TFM, since the large number of hidden variables, even when used with sparsity priors, does not enforce sufficient data faithfulness. Instead, we find that the simplest-possible Bayesian models with global priors yields robust results that can be interpreted as optimal L2 regularization. We study two variants of this algorithm: Bayesian L2 regularization (BL2), where the magnitude of the noise in the displacement data must be measured separately, and Advanced Bayesian L2 regularization (ABL2), which requires no extra input. We test the Bayesian methods using artificial data and real experimental data. Our results suggest that BL2 is not only very robust, but also superior to classical L2 regularization when measurement noise is large. Most importantly, BL2 automatically determines the degree of regularization, which removes subjectiveness from the result. This advance is particularly relevant for in-detail comparison of cells in different conditions, where the varying signal-to-noise ratio previously made an unambiguous comparison challenging.
\section*{Methods}
\subsection*{Regularization}
A common heuristic approach to solve Eq.~\eqref{eq:2} is regularization of the solution through additional constraints. Here, not only the residual of ($\mathbf{u}-\mathbf{Mf}$) is minimized in a least-squares sense, but also the magnitude of the solution is penalized through its p-norm denoted by $\|\mathbf{f}\|_{p}$. The trade-off between minimization of the residual and minimization of the solution norm is determined by fixed regularization parameters, $\lambda_1$ and $\lambda_2$, leading to a minimization problem of the type
\begin{equation}
\hat{\mathbf{f}}=\underset{\mathbf{f}} {\text{argmin}} \big[\|\mathbf{M}\mathbf{f}- \mathbf{u}\|^2_2+\lambda_1\|\mathbf{R}_1\mathbf{f}\|_{1}+\lambda_2\|\mathbf{R}_2\mathbf{f}\|^2_{2}\big].
\label{eq:3}
\end{equation}
The two norms are explicitly written as $||\mathbf{x}\|_1 = \sum_{k} |x_k|$ and $||\mathbf{x}\|^2_2 = \sum_{k} x^2_k$. $\mathbf{R}_1$ and $\mathbf{R}_2$ are functions that are to be defined, e.g., as the unit matrix $\mathbf{I}$. Of the large number of existing regularization strategies, we will focus on the following:

\begin{itemize}
\item L2 regularization, employing an $L2$-norm with $\lambda_2>0$ and $\lambda_1=0$ to penalize traction magnitude through $\mathbf{R}_2 = \mathbf{I}$ is currently the most common technique used for TFM.\cite{schwarz2002calculation,sabass2008high,colin2016super} L2 regularization is also known as ridge regression or Tikhonov regularization\cite{tikhonov1977solutions} and this method efficiently produces a continuous and smooth reconstructed traction field. This approach conveys a high level of robustness for the inversion problem in real space and also in Fourier space. See supplementary information for our implementation in real space.\cite{hansen2007regularization}

\item L1 regularization, also called Lasso,\cite{tibshirani1996regression} is realized through setting $\lambda_2=0$, $\lambda_1>0$, and $\mathbf{R}_1 = \mathbf{I}$. With L1 regularization, small values of the reconstructed signal are efficiently set to zero. L1 regularization is therefore frequently used in the the field of compressive sensing (CS),\cite{candes2006robust} where the underlying assumption is that the signal can be represented in a sparse form where all but a few components of the signal vanish. Recently, the technique has found use for TFM\cite{han2015traction,brask2015compressed,sune2016l1,sune2017super} and it is appropriate for traction fields containing few, sparsely located traction hotspots. In this case, it has been found that L1 regularization improves the ability to distinguish different traction hotspots. 

\item 
%\textcolor{red}{The na\"ive elastic net regularization combines L1- and L2 regularization and both parameters through $\lambda_1>0$, $\lambda_2>0$, and $\mathbf{R}_{1,2} = \mathbf{I}$ in Eq.~\eqref{eq:3}. According to ref.~\cite{zou2005regularization}, the reconstruction of elastic net is estimated as $\hat{\mathbf{f}}(\text{elastic net})=(1+\lambda_2)\hat{\mathbf{f}}(\text{na\"ive elastic net})$ and the elastic net (EN) regularization had been proofed as the following equation
%\begin{equation}
%\hat{\mathbf{f}}=\underset{\mathbf{f}} {\text{argmin}} \bigg[\mathbf{f^T}\bigg( \frac{\mathbf{M^T}\mathbf{M}+\lambda_2\mathbf{I}}{1+\lambda_2}\bigg)\mathbf{f}-2\mathbf{u^T}\mathbf{M}\mathbf{f}+\lambda_1\|\mathbf{f}\|_{1}\bigg].
%\end{equation}
% }
The elastic net regularization combines L1- and L2 regularization and both parameters through $\lambda_1>0$, $\lambda_2>0$, and $\mathbf{R}_1=\mathbf{R}_2 = \mathbf{I}$ in Eq.~\eqref{eq:3}. Note that the actual implementation also involves a rescaling of the variables as suggested in Ref.~\cite{zou2005regularization}. This regularization scheme is known to have a better accuracy compared to L1 and L2 regularization if the coefficient matrix has many, correlated entries.\cite{zou2005regularization} EN regularization is well established for a wide variety of applications, most notably the analysis of genetic data,\cite{hannum2013genome,sunagawa2015structure,horlbeck2016compact,reddy2017genetic} but has to date not been used for TFM. For our implementation of the EN we employ the popular convex optimization solver CVX, see supplementary information.\cite{cvx,gb08}

\item Proximal gradient methods are an alternative approach to the optimization problems arising from the non-differentiable target functions in L1~regularization (PGL) and the EN~regularization (PGEN).\cite{combettes2011proximal,parikh2014proximal} Here, the penalty terms are chosen to be wavelet-transforms written as $\mathbf{R}_1\mathbf{f} = 2\sum_l \langle\mathbf{f},\Psi_l\rangle$ and $\mathbf{R}_2\mathbf{f} = \sum_l \langle\mathbf{f},\Psi_l\rangle$, where the $\Psi_l$ constitute an orthonormal Wavelet basis.\cite{donoho1994ideal,donoho1995wavelet} The optimization problem is solved through iterative soft thresholding, where the regularization parameters control the threshold below which the wavelet-coefficients are set to zero. Proximal gradient methods are widely applied for image inpainting, which is the process of reconstructing lost or deteriorated parts of images.\cite{figueiredo2003algorithm,beck2009fast,mosci2010solving,fadili2011total,peyre2011numerical} Therefore, these methods may be useful for TFM where traction images are reconstructed from undersampled displacement data. Details regarding our implementation of is given in the supplementary information. 
\end{itemize}
These schemes have in common that they require the choice of one or two regularization parameters. Selecting the optimal regularization parameters is often a non-trivial problem. For L2- and L1-regularization, one can use the so-called L-curve criterion~\cite{hansen2007regularization} to find regularization parameters that provide a tradeoff between minimization of residual from the inverse problem and the regularization penalty.\cite{schwarz2002calculation,sabass2008high,han2015traction,colin2016super, legant2013multidimensional,kulkarni2018traction} Usually, the regularization parameter is assumed to be located at the inflection point of a curve described by the norm of the residual versus the norm of the solution in double-logarithmic axes. However, the L-curve criterion is of limited use for real data, since the inflection point does not always exist. Alternatively, multiple inflection points can appear, and the points are hard to localize precisely on the employed logarithmic scales. Moreover, the L-curve criterion does not behave consistently in the asymptotic limit of large system sizes or when the data is strongly corrupted by noise.\cite{hansen1999curve,hanke1996limitations} Hence, in practice, regularization parameters are often chosen by visual inspection of the resulting traction field. This procedure lacks objectivity and significantly biases any conclusions drawn from later analysis of the traction forces. Note that this problem is not specific to regularization, but the issue of distinguishing between noise and ``real'' signal appears generally with any type of method if the data is processed in any way to reduce noise.
\subsection*{Bayesian approaches for traction reconstruction}
The discrete inverse problem $\mathbf{u}=\mathbf{M}\mathbf{f} + \mathbf{s}$ is of two-fold statistical nature. First, the traction varies from cell to cell and from image to image. Thus, $\mathbf{f}$ is a sample drawn from a distribution of possible traction values, which we denote by $p(\mathbf{f}|\alpha)$ with an undetermined parameter $\alpha$. The function $p(\mathbf{f}|\alpha)$ describes any prior knowledge about the distribution of traction. For reasons that will become clear below, we will assume that the prior distribution for the $2 n\times 1$ vector $\mathbf{f}$ is a Gaussian
\begin{equation}
 p(\mathbf{f}|\alpha)=\frac{\exp[-\alpha E_\text{f}(\mathbf{f})]}{Z_\text{f}},
 \label{eq:6}
 \end{equation}
where $Z_\text{f}=(2\pi/\alpha)^n$ and $E_\text{f}= \mathbf{f^T} \mathbf{f}/2$. The second source of randomness is the acquisition noise $\mathbf{s}$. Typically, $\mathbf{s}$ is assumed to be drawn from a zero-mean Gaussian with unknown variance $1/\beta$.~\cite{mackay1992bayesian,molina1999bayesian,suyu2006bayesian,ghosh2015bayesian} In the language of Bayesian statistics, the probability distribution $p(\mathbf{u}|\mathbf{f},\beta)$ is called the likelihood function and determines the probability to to measure a particular vector $\mathbf{u}$ given a traction vector $\mathbf{f}$. Since the noise is Gaussian, the likelihood function is 
 \begin{equation}
 p(\mathbf{u}|\mathbf{f},\beta)=\frac{\exp[-\beta E_\text{u}(\mathbf{u}|\mathbf{f})]}{Z_\text{u}},
 \label{eq:5}
 \end{equation}
where $E_\text{u}(\mathbf{u}|\mathbf{f})=\mathbf{(Mf-u)^T}\mathbf{(Mf-u)}/2$ and $Z_\text{u}=(2\pi/\beta)^m$. $m$ is, as above, the number of two-dimensional displacements. The likelihood function $p(\mathbf{u}|\mathbf{f},\beta)$ describes a situation that is exactly the reverse of the experimental situation, where we are looking for the probability of having $\mathbf{f}$ given measurements $\mathbf{u}$. 
%\textcolor{red}{This situation is described by the posterior distribution $p(\mathbf{f}|\mathbf{u})$, assumed as Gaussian distribution~\cite{mackay1992bayesian,suyu2006bayesian}, and can be related to the likelihood and prior via Bayes' rule 
%\begin{equation}
% P(\mathbf{f}|\mathbf{u},\alpha,\beta)=\frac{p(\mathbf{u}|\mathbf{f},\beta)p(\mathbf{f}|\alpha)}{p(\mathbf{u}|\alpha,\beta)}=\frac{\exp[-K(\mathbf{f})]}{Z_{\text{K}}},
%\label{eq:4}
%\end{equation}
%where $K(\mathbf{f}) = \alpha E_\text{f}(\mathbf{f})+\beta E_\text{u}(\mathbf{f})$ and $Z_{\text{K}}=\int \mathrm{d}^{2n}\mathbf{f}\exp[-K(\mathbf{f})]$.} 
This situation is described by the posterior distribution $p(\mathbf{f}|\mathbf{u})$ and can be related to the likelihood via Bayes' rule     
\begin{equation}
P(\mathbf{f}|\mathbf{u},\alpha,\beta)=\frac{p(\mathbf{u}|\mathbf{f},\beta)p(\mathbf{f}|\alpha)}{p(\mathbf{u}|\alpha,\beta)}=\frac{\exp[-\beta E_\text{u}(\mathbf{u}|\mathbf{f})]}{Z_\text{u}}\frac{p(\mathbf{f}|\alpha)}{p(\mathbf{u}|\alpha,\beta)}.
\label{eq:4}
\end{equation}
Here, the marginal likelihood $p(\mathbf{u}|\alpha,\beta)$ is the overall probability of finding the displacements $\mathbf{u}$ when the traction distributions are integrated out. Thus, $p(\mathbf{u}|\alpha,\beta)$ is also called evidence for the model with $\{\alpha,\beta,\mathbf{u}\}$. 

Assuming that $\alpha$ and $\beta$ are known constants, one can maximize the posterior probability $P(\mathbf{f}|\mathbf{u},\alpha,\beta)$ with respect to $\mathbf{f}$. The resulting solution then satisfies $\mathbf{f}_\text{MP}=\text{argmin}_{\mathbf{f}} \big[\beta\|\mathbf{M}\mathbf{f}- \mathbf{u}\|^2_2+\alpha\|\mathbf{f}\|^2_2\big]$, which is exactly the formula employed for L2 regularization, Eq.~(\ref{eq:3}), if the parameters $\alpha$ and $\beta$ are related to the L2 regularization parameter as $\lambda=\alpha/\beta$.\cite{plotnikov2014high} Thus, our choice of a Gaussian prior is justified if we intend to perform an L2 regularization. Other popular choices for replacing Eq.~(\ref{eq:5}) as prior are the Laplace distribution $p(\mathbf{f}|\theta)=(\theta/2)\exp(-\theta/2 \|\mathbf{f}\|_1)$,\cite{babacan2010bayesian} and a product of a Gaussian and a Laplace distribution.\cite{li2010bayesian} Using these priors, we would have found the formulas corresponding to L1 regularization and EN regularization, respectively. Thus, regularization is equivalent to maximizing the posterior probability of a measurement assuming fixed, known parameters of the prior distributions.

However, $\alpha$ and $\beta$ can also be treated as variables whose values can be determined by maximizing their probability $p(\alpha,\beta|\mathbf{u}) = p(\mathbf{u}|\alpha,\beta)p(\alpha,\beta)/p(\mathbf{u})\sim p(\mathbf{u}|\alpha,\beta)$, where we assume a uniform prior $p(\alpha,\beta)$ and we can omit the marginal probability since it plays no role for the optimization. To calculate the evidence $p(\mathbf{u}|\alpha,\beta)$, we need to integrate out $\mathbf{f}$ in the posterior given in Eq.~(\ref{eq:4}). Due to the Gaussian probabilities, this integration can be done conveniently by expanding the integrand around the most probable value $\mathbf{f}_\text{MP}$. On defining $K(\mathbf{f}) \equiv \alpha E_\text{f}(\mathbf{f})+\beta E_\text{u}(\mathbf{f})$ and its Hessian, $\mathbf{A}\equiv\nabla\nabla \mathbf{K}(\mathbf{f})$, we expand as $K(\mathbf{f}) \approx K(\mathbf{f}_{\text{MP}}) + (\mathbf{f}-\mathbf{f}_\text{MP})^T\mathbf{A}(\mathbf{f}-\mathbf{f}_\text{MP})/2$. Thus, the evidence becomes
\begin{align}
p(\mathbf{u}|\alpha,\beta) =\int_{\mathbf{f}} p(\mathbf{u}|\mathbf{f},\beta)p(\mathbf{f}|\alpha)\mathrm{d}^{2n}\mathbf{f}= \frac{1}{Z_{\text{u}}Z_{\text{f}}}\int_{\mathbf{f}} \exp[-K(\mathbf{f})]\,\mathrm{d}^{2n}\mathbf{f} = 
\frac{(2\pi)^n(\det\mathbf{A})^{-1/2}}{Z_{\text{u}}Z_{\text{f}}}\exp[-\mathbf{K}(\mathbf{f}_\text{MP})].
\end{align}
%\textcolor{red}{According to $K(\mathbf{f})$ and its Hessian, $\mathbf{A}\equiv\nabla\nabla \mathbf{K}(\mathbf{f})$, we expand as $K(\mathbf{f}) \approx K(\mathbf{f}_{\text{MP}}) + (\mathbf{f}-\mathbf{f}_\text{MP})^T\mathbf{A}(\mathbf{f}-\mathbf{f}_\text{MP})/2$. Thus, the evidence becomes
%\begin{align}
%p(\mathbf{u}|\alpha,\beta) =\frac{p(\mathbf{u}|\mathbf{f},\beta)p(\mathbf{f}|\alpha)}{P(\mathbf{f}|\mathbf{u},\alpha,\beta)}=\frac{Z_{\text{K}}}{Z_{\text{u}} Z_{\text{f}}}= \frac{1}{Z_{\text{u}}Z_{\text{f}}}\int_{\mathbf{f}} \exp[-K(\mathbf{f})]\,\mathrm{d}^{2n}\mathbf{f} = 
%\frac{(2\pi)^n(\det\mathbf{A})^{-1/2}}{Z_{\text{u}}Z_{\text{f}}}\exp[-\mathbf{K}(\mathbf{f}_\text{MP})].
%\end{align}
%}
Taking the logarithm yields the final result 
 \begin{equation}
 \log p(\mathbf{u}|\alpha,\beta)=-\alpha E_\text{f}(\mathbf{f}_\text{MP})-\beta E_\text{u}(\mathbf{f}_\text{MP})-\frac{1}{2}\log(\det\mathbf{A})+n\log\alpha+m\log\beta-m\log(2\pi).
 \label{eq:7}
 \end{equation}
The right hand side of this equation is a typical example for target functions employed in Bayesian analysis, for example in the context of data fitting.\cite{mackay1992bayesian} For TFM, numerical calculation of $\log(\det\mathbf{A})$ requires some care. We employ here a Cholesky decomposition of the positive matrix $\mathbf{A}=\mathbf{L}\mathbf{L}^T$ yielding $\log(\det\mathbf{A})= \log(\det(\mathbf{LL^T}))=2\log\Pi_iL_{ii}=2\Sigma_i\log(L_{ii})$.

The logarithmic evidence, Eq.~(\ref{eq:7}), assumes a maximum at those parameters $\hat{\alpha}$ and $\hat{\beta}$ that are most likely associated with the measurement $\mathbf{u}$. The implicit equations resulting from maximizing Eq.~(\ref{eq:7}) read\cite{mackay1992bayesian} $2\hat{\alpha} E_\text{f}^{\text{MP}}=2n-\hat{\alpha} \text{Tr}\mathbf{A}^{-1}$ and $2\hat{\beta} E_\text{u}=2m-2n+\hat{\alpha} \text{Tr}\mathbf{A}^{-1}$. Since $\mathbf{f}_\text{MP}$ and $\mathbf{A}$ depend on $\alpha$ and $\beta$, these equations need to be solved numerically. Once $\hat{\alpha}$ and $\hat{\beta}$ are determined, the equivalent optimal L2 regularization parameter follows as $\hat{\lambda}=\hat{\alpha}/\hat{\beta}$.

We employ two approaches for determining the numerical values of $\alpha$ and $\beta$. In the first approach, called Bayesian L2 regularization (BL2), we estimate the inverse noise variance $\beta$ directly from the data calculating the variance of the measured displacements in spatial regions that are very far away from any cell. Thus, in BL2 only $\alpha$ is determined through maximization of Eq.~(\ref{eq:7}). In the second approach, termed advanced Bayesian L2 regularization (ABL2), we solve directly for $\alpha$ and $\beta$, which result in an increased computational cost. For both approaches, it is imperative to standardize the data to adjust its spread in different dimensions. For a displacement vector $\mathbf{u}$ of length $2m$, we first subtract the mean $\tilde{\mathbf{u}}=\mathbf{u}-\mathbf{1}_{2m}\bar{u}$ with $\bar{u}=\sum_{i=1}^{2m}u_i/(2m)$. Next, we calculate the mean and standard deviation for all columns of the matrix $\mathbf{M}$ as $\bar{M}_j=1/(2m)\sum_{i=1}^{2m}M_{ij}$ and $\omega_j=(\sum_{i=1}^{2m}(M_{ij}-\bar{M}_j)^2/(2m-1))^{1/2}$. Thus, we can define a problem matrix where each column is normalized by its spread $\tilde{M}_{ij}= (M_{ij} - \bar{M}_j)/\omega_j$. The standardized problem therefore reads $\tilde{u}_i = \tilde{M}_{ij}\tilde{f}_j$, which yields $f_i=\tilde{f}_i/\omega_i$. 
 
\subsection*{Generation of artificial test data}
To quantitatively compare the performance of different reconstruction methods, we require artificial data with exactly known traction force and displacements. The process of generating this data is shown in Fig.~\ref{fig:2}(a,i)-(a,iv) and involves prescribing traction force magnitude and direction in distributed circular areas, analytical calculation of the resulting displacements,\cite{sabass2008high} sampling displacements at discrete positions and addition of noise, and finally the reverse traction reconstruction. Supplementary S5 provides further details on the involved analytical calculations. Throughout the article, artificial test data is generated for gel substrates with a Young modulus of $E=10\,\,\mathrm{kPa}$ and a Poisson ratio of $\nu=0.3$. The size of the image plane is arbitrary, but fixed to $25\,\mu\rm{m}\,\times\,25\,\mu\,\rm{m}$ and involves 9 or 15 circular traction spots. For these fixed geometries we vary the traction magnitude, density of displacements, and the noise level.

\subsection*{Evaluation metrics for assessing the quality of traction reconstruction}
To evaluate the quality of the reconstructed traction, we introduce four different error measures comparing reconstructed traction and known original traction. For this purpose, traction at every grid node is written as a two-dimensional vector $\mathbf{t} = \{t_x,t_y\}$. Real traction and reconstructed traction are discriminated by superscripts as $\mathbf{t}^{\text{real}}$ and $\mathbf{t}^{\text{recon}}$.
The error measures are calculated by discriminating traction inside and outside of $N_i$ circular traction patches in a test sample.

\begin{itemize}
\item The Deviation of Traction Magnitude at Adhesions (DTMA)\cite{sabass2008high} is defined as 
\begin{equation}
\text{DTMA}=\frac{1}{N_i}\sum_{i}\frac{\text{mean}_j\left(\|\mathbf{t}^{\text{recon}}_{j,i}\|_2-\|\mathbf{t}^{\text{real}}_{j,i}\|_2\right)}{\text{mean}_j\left(\|\mathbf{t}^{\text{real}}_{j,i}\|_2\right)},
\label{DTMA}
\end{equation}
where $N_i$ is the number of circular traction patches and the index $i$ runs over all patches. The index $j$ runs over all traction vectors in one patch. A DTMA of 0 represents a perfect average traction recovery and a negative or positive value implies underestimation or overestimation, respectively.

\item The Deviation of Traction Magnitude in the Background (DTMB) is the normalized difference between the reconstructed and real traction magnitude outside the circular patches
\begin{equation}
\text{DTMB}=\frac{\text{mean}_k\left(\|\mathbf{t}^{\text{recon}}_{k}\|_2-\|\mathbf{t}^{\text{real}}_{k}\|_2\right)}{\frac{1}{N_i}\sum_{i}\text{mean}_j\left(\|\mathbf{t}^{\text{real}}_{j,i}\|_2\right)},
\label{DTMB}
\end{equation}
where the index $k$ runs over all traction vectors outside the patches. A DTMB with a magnitude much smaller than unity implies low background noise in the reconstructed traction.

\item The Signal to Noise Ratio (SNR) for TFM 
\begin{equation}
\text{SNR}=\frac{\frac{1}{N_i}\sum_i \text{mean}_j(\|\mathbf{t}^{\text{recon}}_{j,i}\|_2)}{ \text{std}_k(\mathbf{t}^{\text{recon}}_{k})}.
\label{SNR}
\end{equation}
measures the detectability of a real signal within a noisy background.\cite{holenstein2017high} As before, the index $k$ runs over all traction vectors outside the patches while $j$ is the index of each traction vector in the patch $i$. The value of the SNR runs from 0 to infinity where a SNR that is much larger than unity indicates a good separation between traction and noise. 

\item The Deviation of the traction Maximum at Adhesions (DMA) 
measures how peak-values of the traction over- or underestimate the true value. The quantity is defined as
\begin{equation}
\text{DMA}=\frac{1}{N_A}\sum_i\frac{ \left[\text{max}_j(\|\mathbf{t}^{\text{recon}}_{j,i}\|_2)-\text{max}_j(\|\mathbf{t}^{\text{real}}_{j,i}\|_2)\right]}{\text{max}_j(\|\mathbf{t}^{\text{real}}_{j,i}\|_2)},
\label{DMA}
\end{equation} 
where the maxima of traction magnitude are calculated for each traction patch separately through index $j$. This error measure is particularly important since traction maxima are easy to extract from real experimental data. A DMA of 0 means that the local traction maxima in the reconstruction and in the original data are equal. Positive or negative values of the DMA indicate that the maximum of traction is overestimated or underestimated. 
\end{itemize}

\subsection*{Experimental procedures}
Primary murine podocytes were isolated and maintained by following previously published protocols.~\cite{schell2013n} In brief, mGFP positive podocytes were isolated from mTom/mGFP*Nphs2Cre reporter mice and subsequent FACS based purification resulted in a primary podocyte culture of highest purity.~\cite{schell2017ferm} Substratum matrices were prepared according to previously established protocols.\cite{plotnikov2014high} Primary podocytes were seeded on polyacrylamide substrates with a Young's modulus of $16$~kPA. Before, gels were functionalized via SULFO-SANPAH based crosslinking of fibronectin (UV light applied for 5 minutes). Cells were cultivated for $12-16$ hours before imaging. Cover slips were transferred into flow chambers and cell removal to image the stress-free gel was achieved with a cell micromanipulator (Eppendorf). 

Heart muscle cells from embryonic rats were freshly prepared as described
previously\cite{HerschMerkel2015}. Cover slides were coated with approximately $70~\mu\rm{m}$ thick
silicone elastomer layer produced from a commercial two-component formulation
(Sylgard $184$, Dow Corning; mixing ratio 50:1 base to crosslinker by weight; cured
overnight at $60^{\circ}\mathrm{C}$). These substrates contained fluorescent beads in their
uppermost layer (FluoSpheres Crimson carboxylate-modified beads;
Invitrogen) and were coated with fibronectin before cell seeding. Details on
preparation and cell culture are published elsewhere\cite{HerschMerkel2015}, calibration of stiffness\cite{CesaMerkel2007} yielded a Young module of 15~kPa and a Poisson ratio of~0.5. Live cell microscopy on spontaneously beating cardiac
myocytes was performed and positions of fluorescent
beads were determined by cross-correlation as described in detail previously.\cite{Merkel2007,HerschMerkel2015}
\section*{Results}
\subsection*{Manual selection of optimal regularization parameters is challenging} 
\begin{figure}[ht]
	\centering
	\includegraphics[width=1\linewidth]{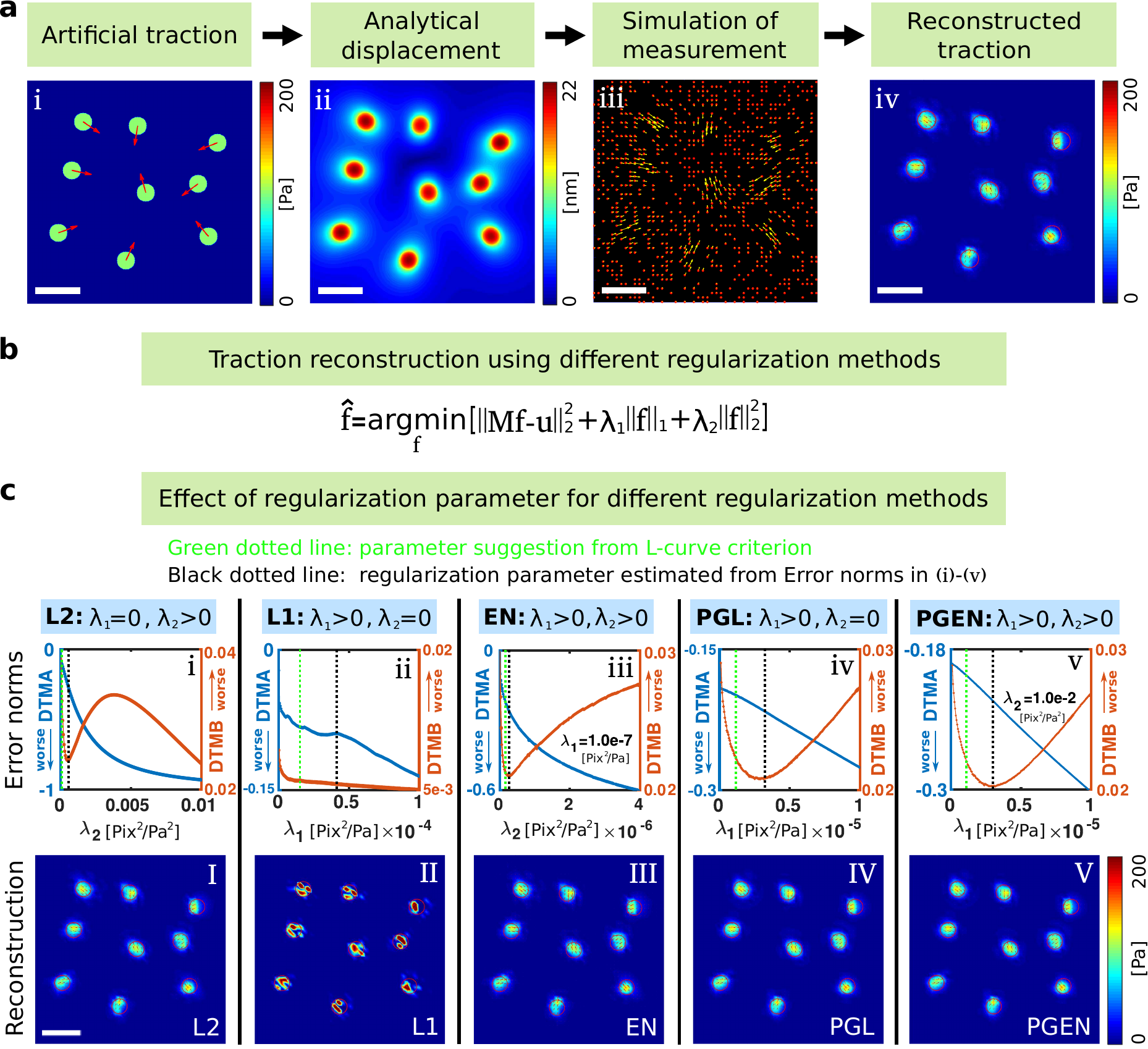}
	\caption{Systematic tests illustrate substantial ambiguity in the choice of regularization parameters. (a)~Schematic of the employed procedure to test the reconstruction methods. (a,i)~Artificial traction pattern consisting of circular spots that uniformly exert a traction of 100~Pa. (a,ii)~Analytical calculation of the gel displacements. (a,iii)~The displacement field is sampled at random positions representing measurements of motion of fiducial markers. (a,iv)~Reconstruction of the traction. (b) Central formula summarizing different regularization approaches. (c)~Dependence of various error measures on the regularization parameters. (b,i)-(b,v)~Error measures defined in Eqns.~\eqref{DTMA}-\eqref{DMA} exhibit various extrema and turning points, making the definition of an optimal parameter challenging.	Note that the minima of the errors do not correspond to values of regularization parameters suggested by the L-curve criterion (Green dotted lines vs. black dotted lines). DTMA: Deviation of traction magnitude at adhesion, DTMB: deviation of traction magnitude in background. (b,I)-(b,V)~Traction fields calculated with regularization parameters that correspond to the error minima at the black dotted lines. Space bar:~$5\,\mu\rm{m}$.
	}
	\label{fig:2}
\end{figure}

The optimal regularization parameters $\lambda_{1/2}$ in Eq.~\eqref{eq:3} are usually unknown. Classical methods for their choice are the L-curve criterion\cite{hansen1999curve,winters2010sparsity} or the generalized cross validation (GCV) for L2 regularization.\cite{golub1979generalized,hansen2007regularization} However, these two methods hardly ever produce the same parameter values and results can differ substantially in the presence of noise, see supporting Fig.~S1. To illustrate the strong effect of regularization on traction reconstruction, we focus on artificial test data where the underlying traction pattern is known. Figure~\ref{fig:2}(a) illustrates the generation of artificial traction fields consisting of circular patches each exerting $100\,\,\mathrm{Pa}$. Fig.~\ref{fig:2}c) demonstrates how variation of the regularization parameters affects the error of traction reconstruction with different methods. Note that the errors exhibit minima for intermediate values of the regularization parameters. For the methods shown in panels i,iv,v of Fig.~\ref{fig:2}(c) (L2 regularization, PGL, PGEN), minima occur in the positive error of the background traction DTMB. In contrast, L1 regularization shown in panel ii of Fig.~\ref{fig:2}(c) exhibits a maximum in the DTMA, indicating a parameter regime where the traction magnitude is not underestimated.

The occurrence of clear minima in the error measures suggests that the corresponding regularization parameter values produce a faithful traction reconstruction. Indeed, employing the values corresponding to the error minima yields traction fields that visually compare well with the original data, see Figs.~\ref{fig:2}a) and \ref{fig:2}d). Note that for L1 regularization, the reconstruction clearly overestimates the maximum traction locally. As shown in the supplementary Fig.~S2(b), the overestimation of the maximum quantified by the DMA can only be reduced through $\sim10$ fold reduction of $\lambda_1$, which however leads to strong background traction and suppression of real traction, see also Fig.~S6. While the minima of the error measures in Fig.~~\ref{fig:2}c) allow to determine a ``best'' regularization for test data, the resulting regularization parameter values deviate from those suggested by the L-curve criterion, see green lines in Fig.~~\ref{fig:2}c). Moreover, the L-curves for these samples are complex and exhibit multiple turning points, illustrating the difficulty in choosing the right regularization parameter in experiments, see supporting~Fig.~S5. 
\subsection*{The elastic net outperforms other regularization methods for traction reconstruction}
\begin{figure}[ht]
	\centering
	\includegraphics[width=1\linewidth]{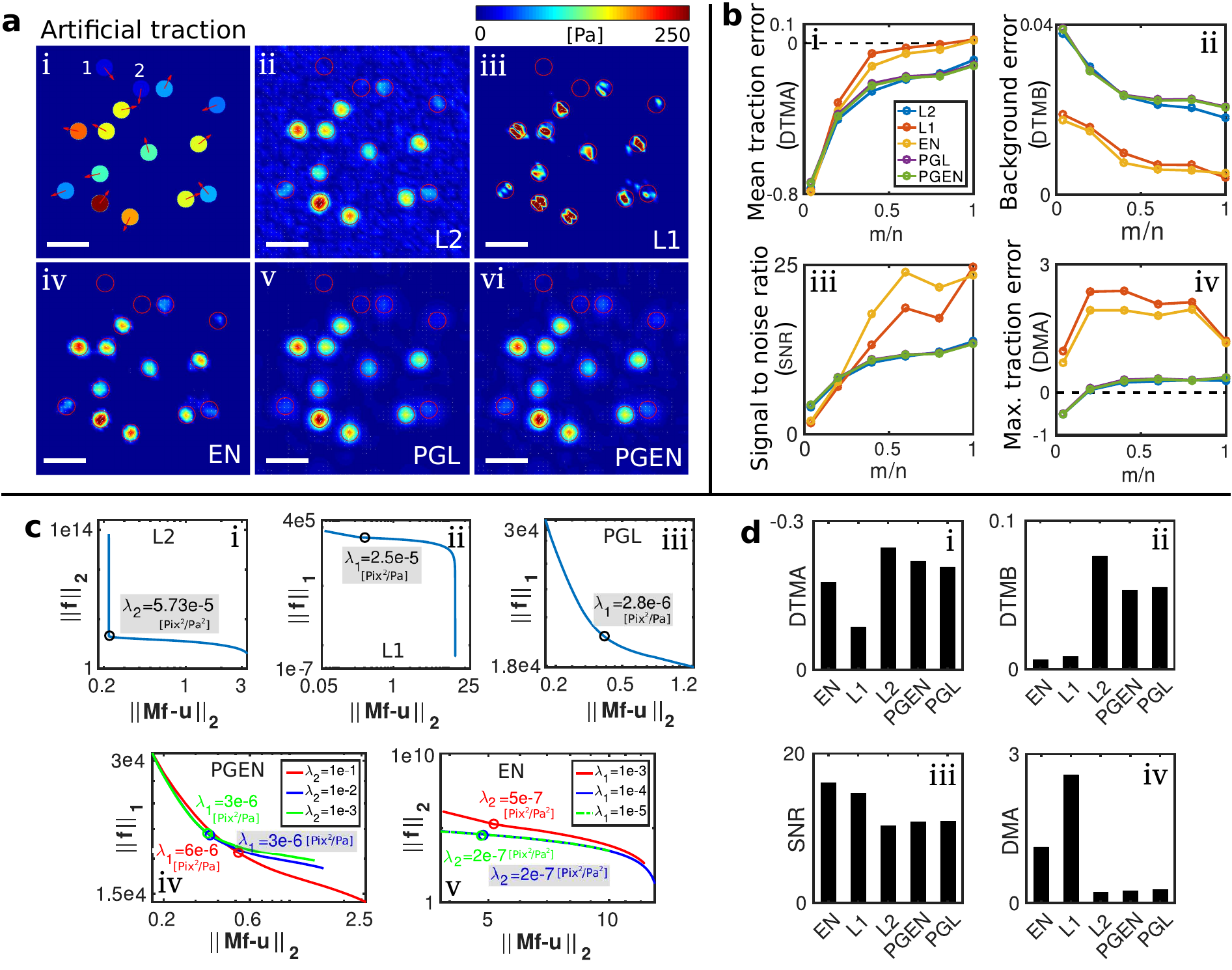}
	\caption{The elastic net (EN) outperforms other reconstruction methods in the presence of noise and when applied to undersampled data.  (a)~Artificial test data with uniform traction spots and 4\% noise in the displacements. Traction maps in (ii-vi) result from different regularization methods: L2-, L1-, EN, PGL and PGEN, respectively. Space bar~$5\,\mu\rm{m}$, displacements are sampled on average every $0.5\,\mu\rm{m}$ (b)~Comparison of errors resulting from undersampled data. Undersampling is realized by reducing the number of displacement vectors $m$. (c) L-curves with chosen regularization parameters (gray boxes) for a data set containing 2\% noise and $m/n=0.4$. (d)~Comparison of errors for the regularization parameters shown in (c). EN regularization shows a favorable tradeoff between error and background signal.}
	\label{fig:3}
\end{figure}

To facilitate quantitative comparison of different reconstruction methods, we employ artificial data consisting of 15 circular traction spots with traction magnitude between 0~Pa and 250~Pa, see Fig.~\ref{fig:3}(a). Gaussian noise with a standard deviation given in percent of the maximal absolute value of the of true displacements is added. The spots have a diameter of 2 $\mu$m and the mesh constant for traction reconstruction is 0.5 $\mu$m.  

Results from different regularization approaches are shown in Fig.~\ref{fig:3}(a). The figure illustrates that L2 regularization can yield realistic estimates for the absolute magnitude of traction on the spots but produces a strong traction background, which may render identification of traction sites difficult. The opposite deficiencies occur for results from L1 regularization. Here, the background is nicely suppressed, which can allow excellent resolution of very small traction spots. However, the peak tractions are significantly overestimated, which can not be mitigated by increasing the regularization parameter, see the supplementary Fig.~S6. Note that the quality of L1 regularization can be improved by using an Iterative Reweighted Least Squares algorithm and the solution from the L2 regularization as an initial guess, see Supplemental Material. The best results are obtained with the EN regularization which combines the advantages of L1- and L2-regularization. Here, we obtain a clean background combined with acceptable accuracy in the absolute traction magnitude on the circular patches. The results from the proximal gradient methods PGL and PGEN qualitatively have a smooth appearance with a level of background traction that is between those of L2 and L1. Fig.~\ref{fig:3}(b) quantifies the described differences between the regularization methods through the error in traction magnitude on the traction spots (DTMA), the error in traction magnitude in the background (DTMB), signal to noise ratio (SNR), and error in maximum traction on the spots (DMA). The supplementary Fig. S9 contains additional plots of these quantities. We find that the reconstruction quality of traction and background improves with increasing number of displacement measurements $m$. Furthermore, EN regularization outperforms other regularization methods with regard to reconstruction accuracy of undersampled data ($m/n<1$). However, the advantage of EN regularization comes at a significantly increased computation time and memory requirement as shown in Table~\ref{tab:1}.
\begin{table}[ht]	
	\centering
	\begin{tabular}{|>{\columncolor[gray]{0.8}}l|c|c|c|c|c|c|c|}		
		\hline
		Reconstruction method & \multicolumn{5}{c|}{Regularization}&  \multicolumn{2}{c|}{Bayesian models} \\\hline
		Name  &L2&L1&EN&PGL&PGEN&BL2&ABL2\\
		\hline
		Simulation time &8 s&75 s&0.8 h&126 s&127 s& 0.1 h &0.5 h\\
		\hline
		Requirement RAM &350 MB&1.98 GB&3.87 GB&101 MB&107 MB& 400 MB & 400 MB\\
		\hline
	\end{tabular}
	\caption{\label{tab:1} Overview of calculation time and RAM requirement for each method. The benchmark results were conducted with a data set consisting of 1000 displacement measurements and a traction field consisting of 2500 entries.}
\end{table}

\subsection*{Bayesian variants of the L2 regularization allow parameter-free traction reconstruction}
\begin{figure}[ht]
	\centering
	\includegraphics[width=1\linewidth]{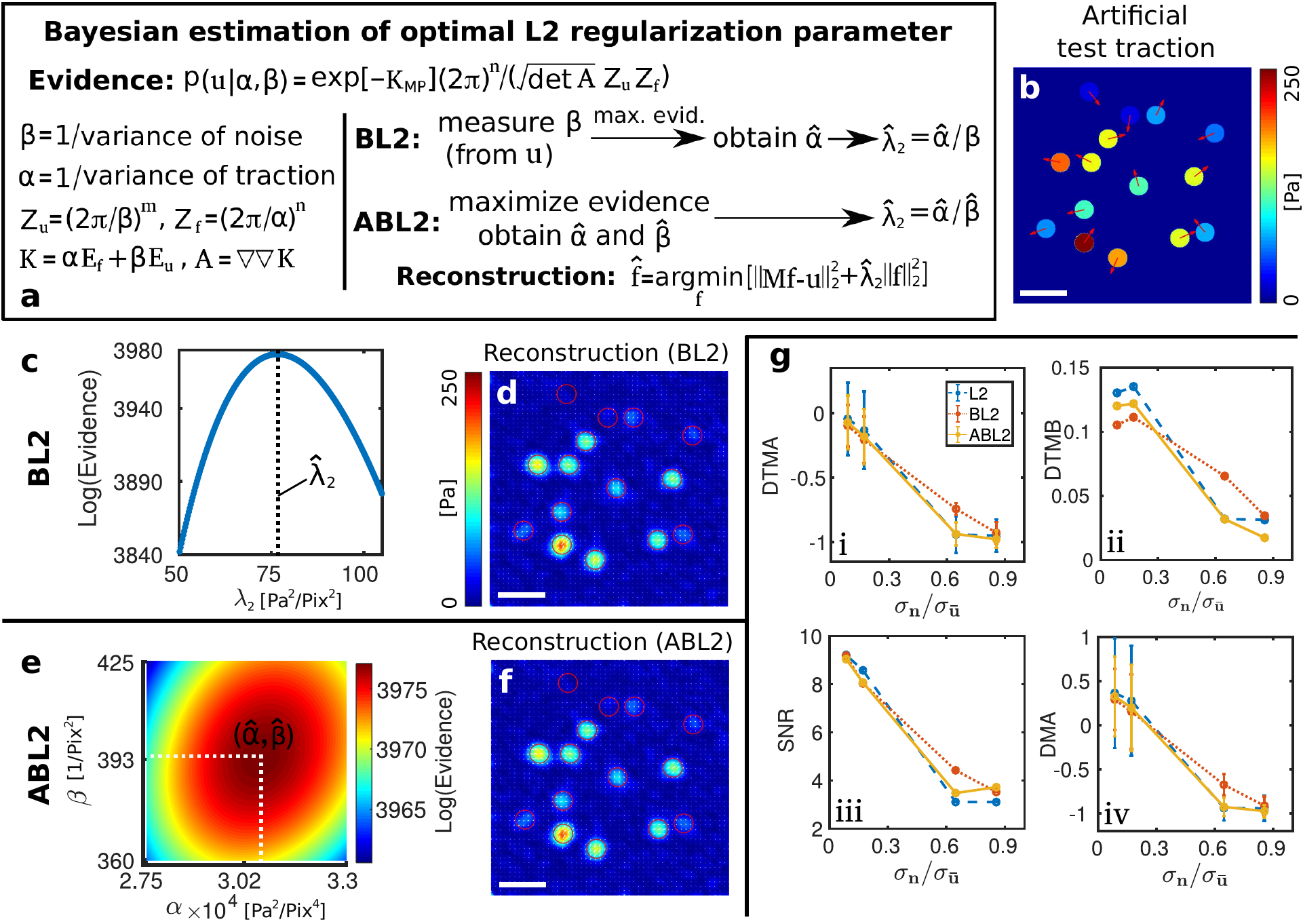}
	\caption{Bayesian L2 regularization (BL2) and Advanced Bayesian L2 regularization (ABL2) are robust methods for automatic, optimal regularization. (a)~Schematic diagram of the procedure employed to infer $\hat{\lambda}_2$ in BL2 and ABL2. BL2 requires the variance of the displacement measurements $1/\beta$, which can be obtained by analyzing displacement noise far away from any cell. ABL2 estimates this noise strength directly from the data. (b)~Artificial test data. For the shown results, 5\% Gaussian noise is added to the displacements. Space bars:~$5\,\mu\mathrm{m}$. (c)~For BL2, the optimal regularization parameter is located at the maximum of a one-dimensional plot of the evidence Eq.~\eqref{eq:7}. (d)~Reconstruction of traction force using BL2. (e)~For ABL2, the optimal regularization parameter is located at the maximum of a two-dimensional plot of the data evidence vs. $\alpha$ and $\beta$. (f)~Reconstruction of traction force using ABL2. (g,i)-(g,iv)~Comparison of reconstruction accuracy for L2, BL2 and AbL2. Different levels of traction forces were applied to change signal-to noise ratio where $\sigma_n$ is the standard deviation of the noise and $\sigma_{\bar{\mathbf{u}}}$ is the standard deviation of the noise-free traction field. Note that BL2 outperforms the other methods for high noise levels.}
	\label{fig:4}
\end{figure}

We next consider the performance of the two Bayesian methods, BL2 and ABL2, that allow automatic choice of the optimal L2 regularization parameter, as schematically shown in Fig.~\ref{fig:4}(a).  Both methods select the optimal regularization parameter by maximizing the logarithmic evidence, Eq.~\eqref{eq:7}. As illustrated in Fig.~\ref{fig:4}(a), the regularization parameter is here deduced from the parameters $\beta$ and $\alpha$, characterizing the distributions of measurement noise and traction respectively. We first employ the same test data as used for Fig. \ref{fig:3}, containing 5\% Gaussian noise in the displacements with $\beta=400 \,\,\mathrm{Pix}^{-2}$. 

With BL2, the log evidence exhibits a clear maximum in a one-dimensional space as seen in Fig.~\ref{fig:4}(c). Figure~\ref{fig:4}(d) shows the reconstructed traction employing the optimal parameter $\hat{\lambda}_2=76.75\,\,\mathrm{Pa}^2/\mathrm{Pix}^2$. Visual comparison of the color-coded traction magnitude in Figs.~\ref{fig:4}(d) and \ref{fig:4}(b) clearly shows that the reconstructed traction has the correct range. 

For ABL2, the evidence is a function of $\beta$ and $\alpha$ as seen in Fig.~\ref{fig:4}(e). Numerical localization of the maximum yields $\hat{\alpha}=3.06e4 \,\,\mathrm{Pa}^{-2}$ and $\hat{\beta}=394\,\,\mathrm{Pix}^{-2}$, which is very close to the known input value of $\beta=400\,\,\mathrm{Pix}^{-2}$. The optimal regularization parameter in this case is thus $\hat{\lambda}_2=\hat{\alpha}/\hat{\beta}=77.66\,\,\mathrm{Pa}^2/\mathrm{Pix}^2$, which agrees well with the estimate from BL2 ($76.75\,\,\mathrm{Pa}^2/\mathrm{Pix}^2$). The resulting traction map is is shown in Fig.~\ref{fig:4}(f) and is very similar to the traction map resulting from BL2 in Fig.~\ref{fig:4}(d). Thus, BL2 and ABL2 yield consistent parameter estimates that produce traction reconstruction of good accuracy. See supplementary Fig.~S9 for a comparison of the Bayesian methods with non-Bayesian approaches.

As with other regularization approaches, quality of reconstruction strongly depends on the present noise. When the magnitude of the noise is comparable to the magnitude of the displacements caused by the traction ($\sigma_{\mathbf{n}}/\sigma_{\bar{\mathbf{u}}}\approx1$), little information can be recovered. For instance, the small circular traction sites labeled 1 and 2 in Fig.~\ref{fig:3}(a) are almost impossible to detect in the presence of $5\,\%$ noise, but can be reconstructed in the noise-free case, see the supplementary Fig.~S7. 
To quantitatively assess the fidelity of reconstruction with small traction forces, we employ a constant $5\,\%$ but scale the tractions to mean values of (12~Pa, 16~Pa, 60~Pa, and 120~Pa). The resulting relative strength of noise and displacements is quantified through the ratio of standard deviations $\sigma_{\mathbf{n}}/\sigma_{\bar{\mathbf{u}}}$, which is plotted against our reconstruction quality measures in Fig.~\ref{fig:4}(g). For comparison, results from manual selection of the regularization parameter using the L-curve criterion are also given. The reconstruction qualities of BL2, ABL2 and L2 are similar when $\sigma_{\mathbf{n}}/\sigma_{\bar{\mathbf{u}}} \ll 1$ (high traction). However, BL2 and ABL2 have an improved signal to ratio SNR compared with the L-Curve approach when $\sigma_{\mathbf{n}}/\sigma_{\bar{\mathbf{u}}}$ approaches unity, see (iii). This is due to the difficulties with the L-curve criterion at high noise. The logarithmic evidence function exhibits in all cases a clear maximum, which enables robust and reliable choice of optimal parameters with BL2 and ABL2. In general, the results from BL2 are however more reliable since the optimization involves here only one parameter. Overall, the tests with artificial data show that these Bayesian methods containing few additional parameters to be determined from the data can resolve the ambiguity associated with manual choice of the regularization parameters over a wide range of signal strengths $\sigma_{\mathbf{n}}/\sigma_{\bar{\mathbf{u}}} < 1$. 

BL2 and ABL2 are based on the simplest structure of a Bayesian model with only one, global prior distribution. One may hypothesize that more complex hierarchies of priors yield an improved traction estimate. For instance, it is possible to prescribe a position-dependent prior for sparse traction patterns through hierarchical Bayesian networks. Such methods require more advanced techniques for sampling of the probability distributions and optimization, such as variational techniques or Markov-chain Monte Carlo methods. We have tested three such algorithms that were originally developed for purposes other than TFM.\cite{hobert1996effect, tipping2003fast, babacan2010bayesian, korobilis2013hierarchical} Results are shown in the supplementary information. However, the tested algorithms all produce highly overestimated, localized traction patterns that sensitively depend on noise. Such errors are likely due to the many free parameters of the models that, in spite of the sparsity constraints, do not favor a faithful data reconstruction. Thus, our tests suggests that these hierarchical network models are not suited for the inverse problem associated with TFM.

\subsection*{Test of methods with experimental data}
\begin{figure}[ht]
	\centering
	\includegraphics[width=1\linewidth]{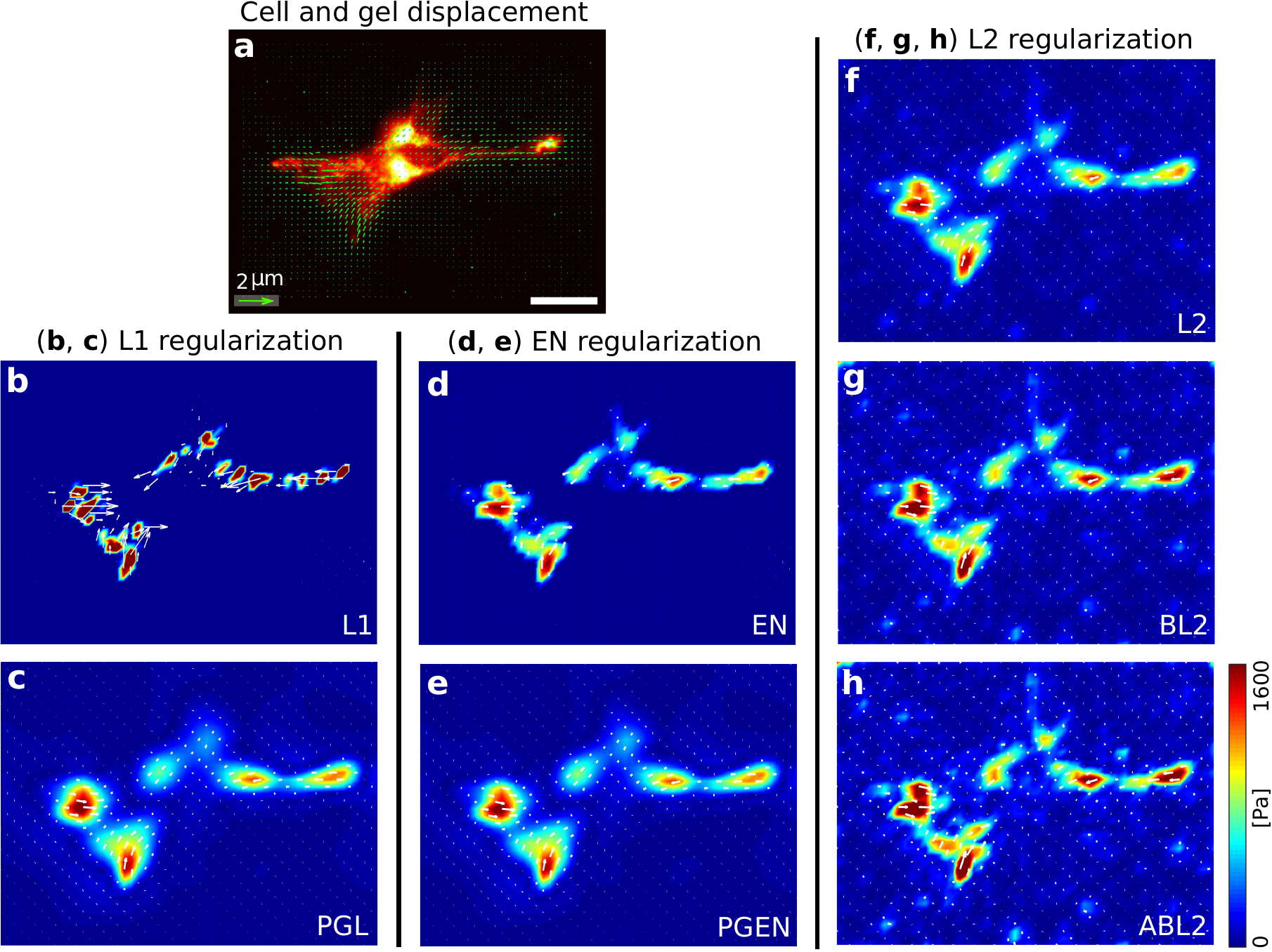}
	\caption{Test of all reconstruction methods using experimental data. (a)~Image of an adherent podocyte with substrate displacements shown as green vectors. (b)-(h)~Reconstructed traction forces using L2, L1, EN, BL2, PGL, PGEN and ABL2, respectively. Reconstruction with L2-type regularization exhibits a comparatively high background noise. L1-regulation shows very high, localized traction. Based on tests with artificial data, we expect that these peaks overestimate the traction. The EN method combines the advantages of L1 and L2 regularization, namely a clean background and localized traction of reasonable magnitude. PGL and PGEN have smooth traction forces at adhesion and background. (g-h)~The Bayesian methods BL2 and ABL2 yield very similar results as the standard L2 regularization without requiring a search for the optimal regularization parameters. For better visibility, only every fourth traction vector is shown. Space bar:~$25\,\mu\rm{m}$.}
	\label{fig:5} 
\end{figure}

To compare the performance of the different methods for real cells, we employ primary mouse podocytes studied with a standard TFM setup
on polyacrylamide gels having a Young module of $\sim16\,\,\mathrm{kPa}$. The deformation field resulting from cellular traction is shown in Fig.~\ref{fig:5}~(a). Using this displacement data, we find that the variance of noise is $\sim0.01\,\rm{pix}^2 = 103.4\,\rm{nm}^2$ in regions that are far away from the cell. The maximum displacement is $0.52\,\mu\mathrm{m}$. Figs.~\ref{fig:5}(b)-(h) show reconstruction results using all methods discussed above. As for artificial data, we find here that the EN regularization results in a very clear background. The traction magnitudes and outlines of adhesion sites are similar to those resulting from regularization with the L2 method. For L1 regularization, traction localizes in sparse regions and has a significantly higher value than for other methods. Proximal gradient methods produce smooth traction profiles as expected from the use of the soft wavelet thresholding. The magnitude of traction measured with PGL and PGEN is close to results of EN and L2.  

Next, we considered the performance of the Bayesian methods. The logarithmic evidence, Eq.~\eqref{eq:7}, calculated with BL2 and ABL2 reveals pronounced maxima, allowing to robustly choose the optimal parameters for the experimental data. See also supplementary Fig.~S11. The resulting values for $\hat{\lambda}_2$ are $30.4\,\,\mathrm{Pa}^2/\mathrm{Pix}^2$ and $24.4\,\,\mathrm{Pa}^2/\mathrm{Pix}^2$ for BL2 and ABL2, respectively, and thus agree reasonably well with each other. Figures~\ref{fig:5}(e),(h) show the traction fields calculated with BL2 and ABL2. These traction fields are visually very similar to the one obtained with standard L2 regularization. However, the L-curve criterion provides a much more uncertain estimate of a regularization parameter due to the difficulty of localizing it on a logarithmic scale, see supplementary Fig.~S12. Note that the regularization parameters obtained from the L-curve criterion can not be directly compared to the parameters resulting from the Bayesian methods due to standardization employed for the latter. Overall, the suggested Bayesian models can eliminate ambiguity in TFM by automatically providing a consistent parameter choice.

\subsection*{Bayesian regularization enables consistent analysis of traction time sequences}
TFM is frequently employed to study dynamical aspects of cell mechanics. Examples include cell migration, cell division, or cytoskeletal reorganization in response to extracellular stimuli. Such processes are usually accompanied with a change in the traction distribution.
As a result, the optimal regularization parameter varies among different images in a time sequence of microscopy data. Additionally, the 
regularization parameter can also change if the degree of noise varies over time, which can be caused for example by stage drift or photo bleaching. In such cases, it is very challenging to perform a consistent, frame-by frame analysis to determine the degree of regularization with conventional methods. Thus, one fixed parameter is commonly employed for the whole time sequence and precision, but also accuracy of traction reconstruction is thereby sacrificed.

\begin{figure}[ht]
	\centering
	\includegraphics[width=1\linewidth]{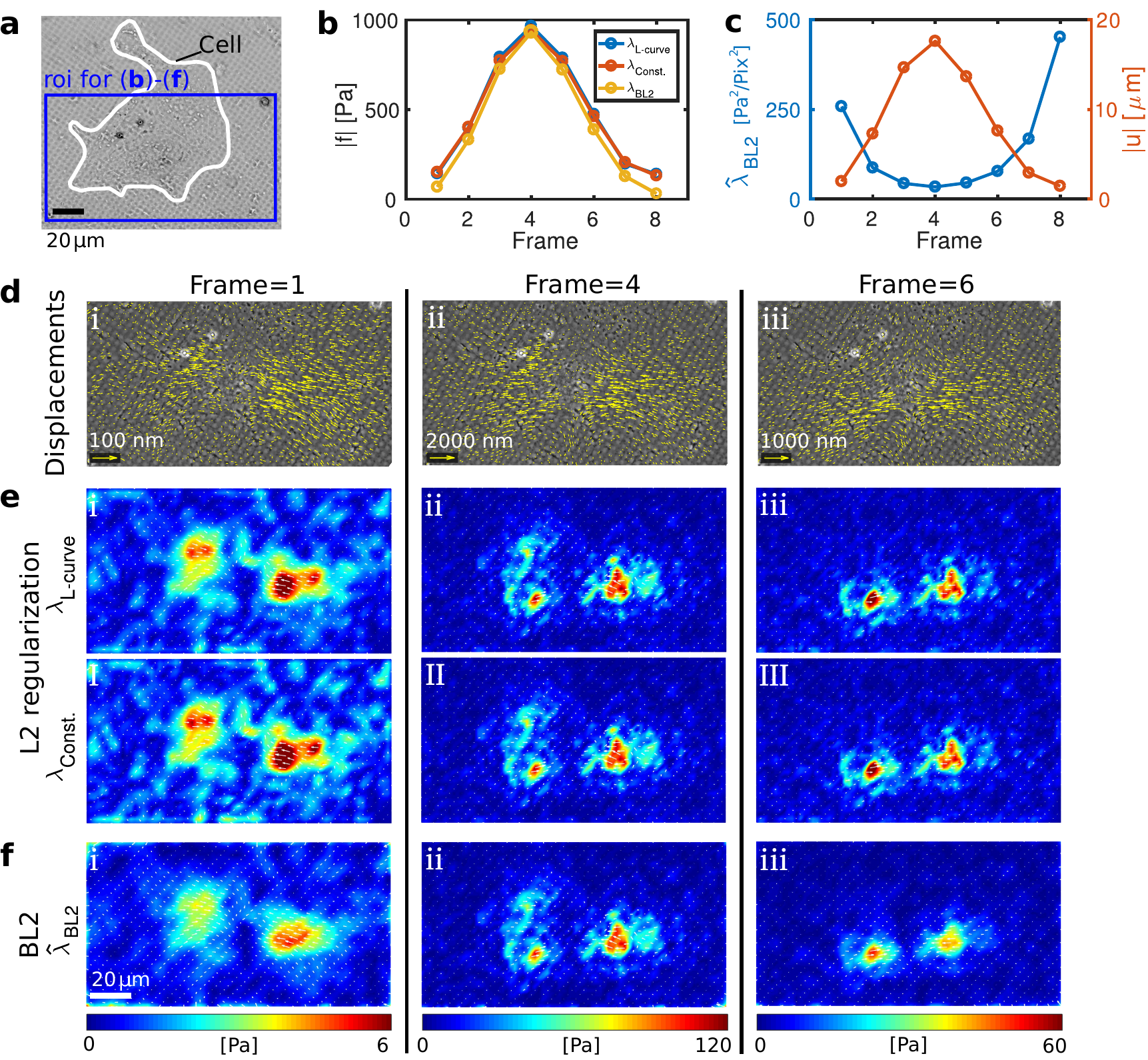}
	\caption{Bayesian L2 regularization robustly adapts to different traction levels allowing quantitative analysis of time series. (a)~Image of a spontaneously beating heart muscle cell on a micropatterned substrate allowing to measure displacements. (b)~Overall norm of traction magitude in successive image frames. The maximum corresponds to one contraction of the heart muscle cell. Traction is calculated with BL2 or, for comparison, via L2 regularization where $\lambda_2$ is either selected manually for every frame using the L-curve criterion or held constant throughout the image sequence. (c)~Optimal regularization parameter suggested by BL2 and the norm displacement field correlate. (d,i)-(d,iii)~Cell images with displacement field at frames $1$, $4$, and $6$. (e,i)-(e,iii)~Snapshots of the traction fields resulting from L2 regularization with a manually chosen parameter $\lambda_\text{L-curve}$ and a constant parameter $\lambda_\text{L-const.}$ in an intermediate range. (f,i)-(f,iii)~Snapshots of the traction fields resulting from BL2. Note the different scaling of displacement and tractions for the different frames. Frame 1 (I) illustrates that BL2 yields a smaller traction magnitude than L2 in the presence of large noise, where the L-curve criterion is hard to employ. As a result, BL2 allows to differentiate real traction from noise outside of the cell. For better visibility, only every fourth traction vector is shown.}
	\label{fig:6}
\end{figure}

To test whether our Bayesian methods can be useful in this situation, we employ TFM data with a spontaneously beating cardiac myocyte, see Fig.~\ref{fig:6}. Due to the large size of the cell, we focus on a region of interest shown in Fig.~\ref{fig:6}(a). The analyzed time sequence corresponds to one cell contraction. Snapshots from frames $1$, $4$, and $6$ are shown for illustration. Figure~\ref{fig:6}(b) shows the overall norm of reconstructed traction where $\lambda_2$ is either chosen according to the L-curve criterion, held at an intermediate, constant value, or automatically determined in BL2. The overall traction magnitudes are similar in frames 2-6 where traction is high. Differences occur, however, in the low-traction regime, where BL2 systematically yields lower values of traction. We expect that the results from BL2 are more trustworthy in this regime since the L-curve criterion yields highly ambiguous values for the regularization parameters, see supplementary Fig.~S13. Figure~\ref{fig:6}(c) shows the overall norm of the gel displacement and the optimal regularization parameter estimated with BL2. The noise variance is small, $\sim0.00003\,\rm{pix}^2$, in regions that are far away from the cell (pixel size $0.2\,\mu\rm{m}$). As expected, $\lambda_{BL2}$ is inversely correlated with the displacement magnitude. 

Figs.~\ref{fig:6}e show snapshots of the resulting traction fields that illustrate again that BL2 produces slightly different results for low traction, most apparent in Figs.~\ref{fig:6}(e)i,I and Fig.~\ref{fig:6}(f),i. Note that the traction field resulting from classical L2 regularization in Figs.~\ref{fig:6}(e),i,I shows a noise background outside of the cell that is almost comparable to the real cellular traction. In contrast, BL2 suppresses this background at the price of an apparently reduced spatial resolution as seen in Fig.~\ref{fig:6}(f),i. However, this provides an objective distinction between real signal and noise, which is what is to be expected from a faithful data reconstruction.

\section*{Discussion}
During the last decade, traction force microscopy has become one of the most popular techniques for studying mechanobiology on the cellular level. The technique has found broad use for studying sub-cellular structures,\cite{balaban2001force,thery2006cell,gardel2008traction,legant2013multidimensional,ray2017anisotropic} single cell dynamics,\cite{chaudhuri2015substrate,valon2017optogenetic,oria2017force,sabass2017force} and collective cellular behavior,\cite{trepat2009physical,mertz2013cadherin,ng2014mapping,steinwachs2015three,lembong2017calcium,MarcqLadoux}, and there are far too many applications to be reviewed here appropriately.

Many, if not all, TFM methods critically rely on some form of noise reduction. Usually, traction is calculated from substrate displacement through the solution of a linear problem involving elastic Green's functions. Here, the effects of noise are not a technical issue relating to the data precision, but connected directly to the structure of the linear problem where even the slightest numerical noise can be amplified to an extent that the true solution is entirely lost. The most immediate approach to deal with noise is to filter the displacement field prior to traction reconstruction. Filtering becomes possible if the solution is calculated in Fourier space because the convolution theorem simplifies the matrix inversion.\cite{butler2002traction} However, filtering the input data generally reduces the spatial resolution and optimal resolution can only be gained if the filter is adapted for each sample. In certain cases, moreover, data filtering is not sufficient to guarantee stability of the solution, for example, if the three-dimensional position of displacements is included. 

A popular alternative strategy for enforcing well-behaved solutions is regularization. With regularization, Fourier-space inversion becomes more robust.~\cite{sabass2008high,kulkarni2018traction} However, regularization is also used for real-space approaches and has been used in conjunction with finite element methods or boundary element methods. Solving the linear problem in real space is generally more demanding, but has the advantage that the spatial sparsity of traction patterns is conserved. For TFM, two regularization methods have to date been used, namely L2 regularization\cite{sabass2008high,plotnikov2014high,holenstein2017high} and L1 regularization\cite{han2015traction,sune2016l1,sune2017super}. These methods each have one regularization parameter that is chosen manually based on heuristics, which introduces a considerable degree of subjectivity in the resulting traction.

In this work, we systematically compare the classical L1- and L2 regularization to three other methods that have, to our knowledge, not yet been employed for TFM. These three regularization methods are the Elastic Net (EN), Proximal Gradient Lasso (PGL) and Proximal Gradient Elastic Net (PGEN). Our tests with artificial data clearly demonstrate that EN regularization outperforms other regularization methods with regard to the reconstruction quality. Here, accurate traction reconstruction is due to a simultaneous suppression of background noise and penalizing of large traction magnitude. In contrast, the proximal gradient methods PGL and PGEN are effective at producing a smoothed traction field, due to the local removal of high-frequency spatial variations. These results obtained with artificial data agree qualitatively with results from tests with experimental data. Here too, L1 and L2 regularization yield overestimated or underestimated traction on adhesion sites. EN again yields a clear background without producing excessively sharp traction peaks at adhesions, see Fig.~\ref{fig:5}. PGL and PGEN yield smooth traction fields and rounded adhesion site contours. 
While our work presents a comprehensive overview of regularization variants in TFM, it does not cover all variants and solution procedures. For example, it has been suggested to use an L1 norm for both the residual and regularization term,\cite{sune2017super} and various other iterative regularization procedures can be tested for TFM in the future.

Next, we ask if Bayesian methods can eliminate the necessity of a manual choice of regularization parameters. Here, the corresponding parameter values are inferred by maximizing their evidence given a fixed class of chosen probability distributions. Using the simplest approach, our prior assumption on the traction forces is that they are drawn from one global Gaussian distribution with an unknown variance $1/\alpha$. The posterior distribution determining the probability of a particular traction field given a measured displacement field is then determined by the parameter $1/\alpha$ and a further parameter $1/\beta$, quantifying the variance of the measurement noise. For fixed values of $\alpha$ and $\beta$, maximization of the posterior distribution corresponds exactly to L2 regularization with $\lambda_2$ given by $\alpha/\beta$. However, the values of $\alpha$ and $\beta$ can also be determined through maximizing their probability conditioned on a given measurement and the chosen probability distributions. Here, this is equivalent to maximizing the evidence for $\mathbf{u}$, given any two parameter values. We refer to the simultaneous determination of both parameters from the evidence as advanced Bayesian L2 regularization (ABL2). In an even simpler approach, we estimate the noise strength directly from the displacement data, leaving only one parameter $\alpha$ to be determined by maximization of the evidence; which we call Bayesian L2 regularization (BL2). These methods represent an automatic optimization of the L2 regularization. Thus the resulting traction field has all the qualitative features of L2 regularization, including the suppression of exceedingly high traction values and a visible background traction. For all our tests, we found that BL2 was a robust method yielding reasonable estimates for traction and regularization parameters. Due to the difficulty in choosing the correct regularization parameter manually, BL2 has substantial advantages over the classical L2 procedure, in particular if the traction is so small that the resulting displacements are comparable to the noise $\sigma_{\mathbf{n}}/\sigma_{\bar{\mathbf{u}}}\approx1$.\\
We mention that we have also tested more elaborate hierarchical Bayesian network algorithms that were originally designed for other purposes than for use with TFM. These include a variational approach termed ``Bayesian compressive sensing using Laplace priors'' (BCSL),\cite{babacan2010bayesian} and Markov chain Monte Carlo methods, for instance the ``Bayesian Lasso''.\cite{korobilis2013hierarchical} In our experience, however, none of these methods could compete with the much simpler Bayesian L2 regularization when applied to TFM problems, see Figs.~S8-S10,S14 in the supplementary information.
 
The advantage of employing Bayesian traction reconstruction is most apparent when cells in different conditions are to be compared. To perform a correct comparison of situations with different traction, different substrate rigidities, etc., it is technically necessary to adapt the regularization parameter for each case. However, the difficulty of finding the corresponding parameters usually makes this impossible, which introduces significant quantitative errors. Bayesian methods present a possible solution to this problem. We have shown here that BL2 produces regularization that varies smoothly and robustly with strong temporal changes in the traction exerted by a cell. Thus, we expect that this method can be of wide use for quantitative studies of cell physiology.
\section*{Acknowledgements}
C.S. acknowledges support by the Berta-Ottenstein Programme, Faculty of Medicine, University of Freiburg, Germany.
\section*{Author contributions statement}
Y.H. implemented the algorithms and analyzed the data; N.H. and R.M. performed the
experiment on heart muscle cells; C.S. and T.H. performed the experiment on podocytes; G.G. discussed data and results; B.S. designed and supervised the project and wrote the main text. All authors participated in writing the manuscript.
\section*{Additional information}
%A free software package for TFM with BL2 and ABL2 is currently being assembled and
%will be made available on the website of the corresponding author.
The authors declare no competing interests. 
%\textbf{Accession codes} (where applicable); 
%\textbf{Competing financial interests} 
%The corresponding author is responsible for submitting a \href{http://www.nature.com/srep/policies/index.html#competing}{competing financial interests statement} on behalf of all authors of the paper. 
%This statement must be included in the submitted article file.

\newpage
%\bibliography{sample}

\newpage

\end{document}